\documentclass[final,5p,times,twocolumn,authoryear]{elsarticle}

\usepackage{graphicx}
\usepackage{subcaption}
\usepackage{amssymb,amsmath}
\usepackage{pdflscape}
\usepackage{adjustbox}
\usepackage{hyperref}
\usepackage{booktabs}
\usepackage{xcolor}
\usepackage{tabularx}
\usepackage{footnote}
\usepackage{soul}
\usepackage{multirow}
\usepackage{multicol}
\usepackage{lineno}
\usepackage{textcomp}
\usepackage{url}
\usepackage[utf8]{inputenc}

\usepackage{amsthm}

\journal{Journal of High Energy Astrophysics}

\begin{document}

\begin{frontmatter}



\title{Short-timescale X-ray variability of the blazar 1ES 1959+650 observed with AstroSat/SXT}


%

\author[label1]{Sikandar Akbar\corref{cor1}}
\ead{darprince46@gmail.com}
\affiliation[label1]{organization={Department of Physics, University of Kashmir},
            city={Srinagar},
            postcode={190006},
            country={India}}

\author[label2]{Zahir Shah\corref{cor2}}
\ead{zahir.shah@manipal.edu}
\affiliation[label2]{organization={Manipal Centre for Natural Sciences, Centre of Excellence, Manipal Academy of Higher Education},
            city={Manipal},
            postcode={576104},
            country={India}}

\cortext[cor1]{Corresponding author}
\cortext[cor2]{Corresponding author}

\begin{abstract}
We analyse the AstroSat Soft X-ray Telescope (SXT) observation of the high-synchrotron-peaked blazar 1ES 1959+650 taken on 2023 September 2, which covers 362\,ks with about 95\,ks of good exposure. We build 0.6--7\,keV light curves in six sub-bands, binned at 5830\,s (the orbital period) and 1000\,s. All sub-bands show excess variance, with fractional variability of 3.7 to 6.5 per cent that increases with energy ($\propto E^{0.32\pm0.07}$). The power spectrum of the total band, fitted with a model that includes the window function of the orbital gaps, is a red-noise power law with slope $-1.80$ (95 per cent range $-2.8$ to $-1.4$). A Lomb-Scargle search finds no significant periodicity (global $p>0.57$) and would have detected, with 90 per cent probability, a sinusoid with an rms amplitude above 1.9 to 3.7 per cent at periods of 3 to 50\,ks. The soft (0.6--1.4\,keV) and hard (1.9--7\,keV) light curves are correlated near zero lag, but only marginally (global $p=0.04$ and 0.07 for power-spectrum slopes of $-1.7$ and $-2.0$), and the lag is $-1^{+6}_{-10}$\,ks, consistent with zero. No pair of sub-bands that includes 0.6--0.8\,keV shows a significant correlation or lag, so an energy-dependent lag of a few ks can be neither confirmed nor excluded. The 0.6--7\,keV light curve contains one hump with rise and decay times of about 0.4 to 0.6\,d, which is only marginally distinguishable from red noise ($p\approx0.05$ to $0.07$). The count rate distribution is consistent with a lognormal distribution, and the hardness ratio does not depend significantly on flux.
\end{abstract}

\begin{keyword}
galaxies: active \sep BL Lacertae objects: individual: 1ES 1959+650 \sep X-rays: galaxies \sep methods: statistical
\end{keyword}

\end{frontmatter}

\section{Introduction}
\label{sec:intro}

Blazars are active galactic nuclei whose emission is dominated by a relativistic jet pointing close to our line of sight \citep{UrryPadovani1995}. In high-synchrotron-peaked (HSP) objects the synchrotron peak lies in the UV to X-ray band, so X-ray observations sample the electrons at the high-energy end of the distribution, close to the peak of the spectral energy distribution. The cooling times of these electrons are of order hours \citep{DasChatterjee2023}, which makes the X-ray band a direct probe of how particles are accelerated and how they lose energy. Open questions include where the emission regions are located and which acceleration process dominates \citep{Markowitz2022}.

Two timing measurements bear on these questions. The first is the shape of the power spectral density (PSD), which for blazars is a red-noise power law that may bend at a characteristic frequency \citep{Chatterjee2018}. The second is the lag between X-ray bands. If electrons are accelerated more slowly than they cool, the harder photons lag the softer ones, and if cooling dominates the softer photons lag \citep{DasChatterjee2023}. Lags between soft and hard X-rays are therefore a way to estimate acceleration and cooling timescales, but they are difficult to measure because they are short compared with the length of typical continuous X-ray light curves, and because gaps in the sampling can produce spurious correlations in red-noise light curves.

AstroSat observations have been used for such measurements. \citet{Markowitz2022} analysed a 4 d AstroSat campaign on Mkn 421 and found that the fractional variability increases with energy across the synchrotron hump. They also found evidence that the 0.6--0.8\,keV band leads the other SXT bands by $4.6\pm2.6$\,ks, but only in the first half of the observation. \citet{DasChatterjee2023} used SXT and LAXPC light curves of Mkn 421 and 1ES 1959+650 at eight epochs and found harder-when-brighter behaviour in all of them, with hard-soft lags that are consistent with zero in some epochs and are a few hours at most in the rest.

1ES 1959+650 is a nearby ($z=0.047$) HSP blazar and has been detected at TeV energies since 1999 \citep{Nishiyama1999}. Its X-ray timing properties have been studied less than those of Mkn 421. \citet{Pandey2017} found no variability timescale in the autocorrelation function of either of two NuSTAR light curves of the source (2014 September 17 and 22, about 20\,ks of good exposure each), and \citet{Bhatta2018} found a correlation between the 3--10 and 10--79\,keV light curves of the same observations that peaks at zero lag, with a peak value of 0.95 in the first and 0.67 in the second. \citet{DasChatterjee2023} measured centroid lags between 0.7--7\,keV (SXT) and 7--20\,keV (LAXPC) of $0.99^{+0.48}_{-0.41}$, $0.09^{+0.53}_{-0.45}$ and $1.98^{+0.36}_{-0.33}$\,hr in three epochs between 2016 and 2017. A flaring state of this source in 2017 was studied in X-rays with AstroSat by \citet{Shah2021} and \citet{Chandra2021}.

Here we analyse an SXT observation of 1ES 1959+650 from 2023 September 2. We measure the band-resolved variability, the correlation and lag between a soft and a hard band, the power spectrum, the periodicities, the shape of the light curve and the flux distribution, with significance estimates that take the orbital gaps of the sampling into account. Section~\ref{sec:data} describes the observation and the construction of the light curves. Section~\ref{sec:results} presents the variability, hardness, lag, power spectrum, periodicity, flux distribution and light curve profile results. We discuss them in Section~\ref{sec:disc} and summarise them in Section~\ref{sec:summary}.

\section{Observation and data reduction}
\label{sec:data}

\subsection{SXT data}

The Soft X-ray Telescope \citep[SXT;][]{Singh2016,Singh2017} on AstroSat is a focusing X-ray telescope with a CCD camera, sensitive between 0.3 and 8\,keV. We used the Level-2 cleaned event files of the observation of 1ES 1959+650 on 2023 September 2, which are produced orbit by orbit, and merged them into a single cleaned event file. The observation spans about 362\,ks from start to end. Earth occultation and passages through the South Atlantic Anomaly leave about 1--3\,ks of good time in each 5.83\,ks orbit (a duty cycle of about 27 per cent), so the effective exposure is about 95\,ks. We extracted source events from a circular region of $15'$ radius centred on the source.

\subsection{Light curves}

We built light curves in the total 0.6--7\,keV band and in six sub-bands: 0.6--0.8, 0.8--1.1, 1.1--1.4, 1.4--1.9, 1.9--3.0 and 3.0--7.0\,keV. Photon energies come from the pulse invariant channels, which are 10\,eV wide. The mean source rate in 0.6--7\,keV is about 5\,counts\,s$^{-1}$.
Count rates are corrected for the fractional exposure of each bin and errors follow the Gehrels approximation \citep{Gehrels1986}. We checked that light curves made with \textsc{xselect} and with our own event-based code agree to better than 0.1 per cent.

We used two binnings. The fractional variability and hardness-ratio checks use bins of 5830\,s, equal to the orbital period, and bins with less than 20 per cent fractional exposure are discarded. The power spectrum, the search for periodicities and the soft-hard cross-correlation use bins of 1000\,s, made by rebinning light curves with 100\,s bins (counts and exposure are summed), and the same exposure cut applies. This gives 131 points in 0.6--7\,keV. Bins of 5830\,s give only about 50 points, which is too few for a power spectrum.

\section{Results}
\label{sec:results}

\subsection{Light curves}

\begin{figure*}
\centering
\includegraphics[width=1.95\columnwidth]{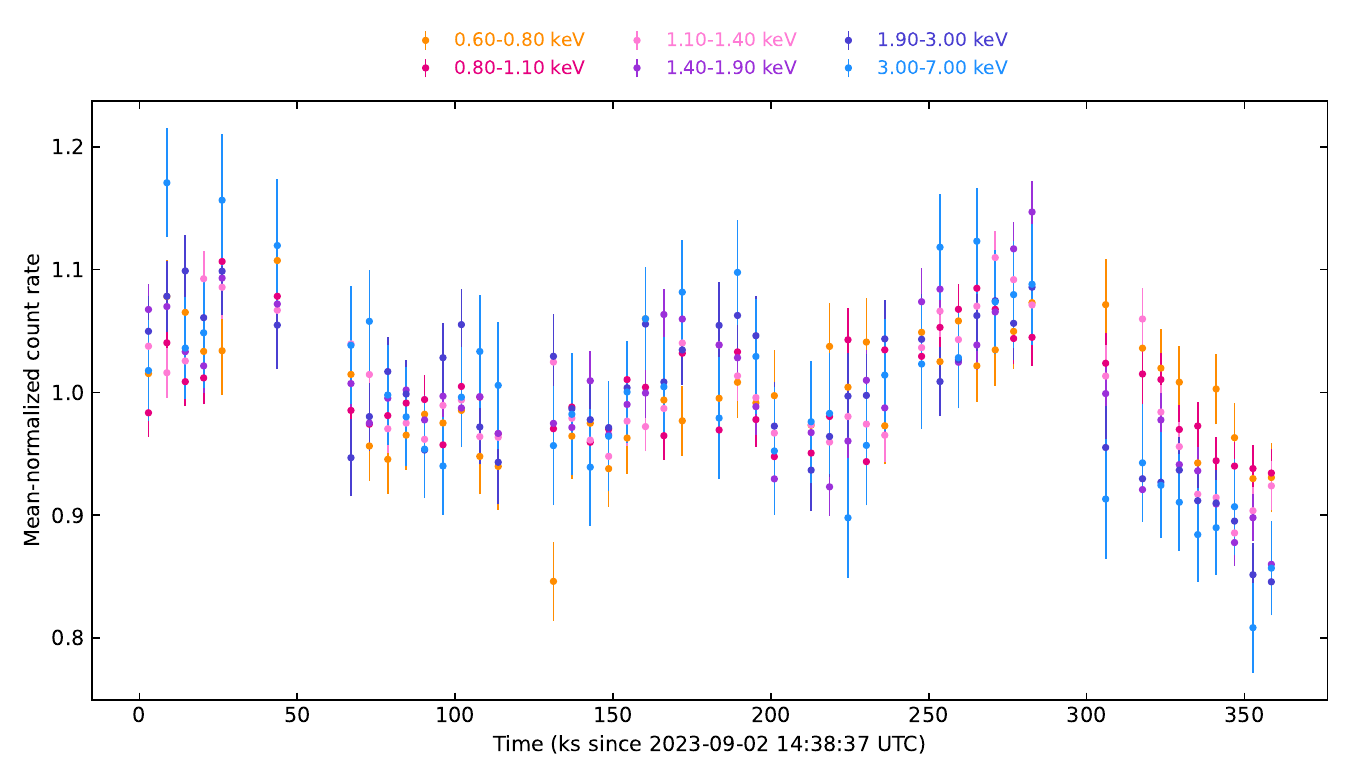}
\caption{SXT light curves of 1ES 1959+650 in the six sub-bands, binned at the orbital period (5830\,s) and normalised to the mean of each band.}
\label{fig:lc}
\end{figure*}

Figure~\ref{fig:lc} shows the six sub-band light curves in 5830\,s bins, each divided by its mean. The 0.6--7\,keV count rate varies between about 4.5 and 5.5\,counts\,s$^{-1}$ around a mean of 5.0. It is high at the start of the observation, lower between about 70 and 200\,ks, highest near 250 to 285\,ks, and falls over the last 60\,ks (Figure~\ref{fig:flare}). By eye the sub-bands move together.

\subsection{Fractional variability}

\begin{figure*}
\centering
\includegraphics[width=1.5\columnwidth]{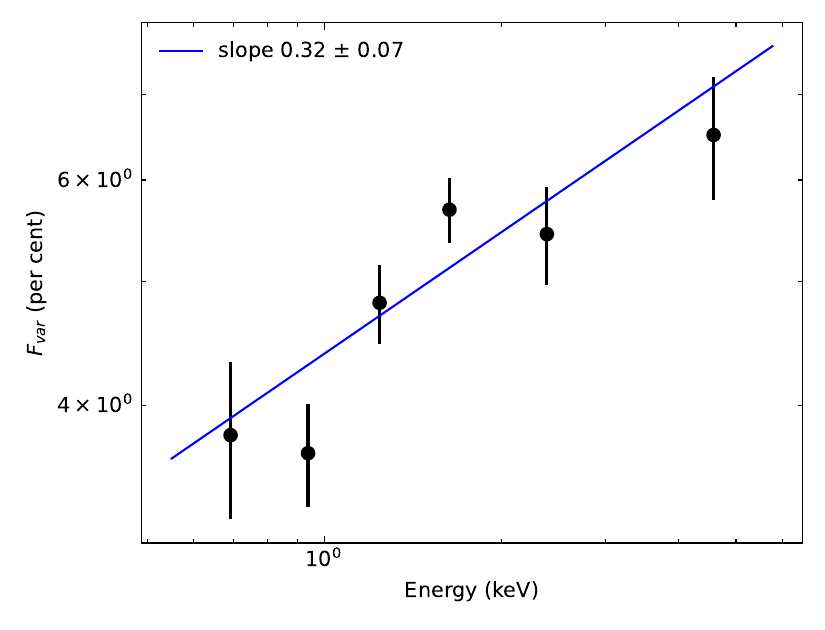}
\caption{Fractional variability as a function of energy.}
\label{fig:fvar}
\end{figure*}

We quantified the variability in each band with the fractional rms amplitude $F_{\rm var}$ and the normalised excess variance \citep{Vaughan2003}. All six sub-bands show excess variance over the full observation, with $F_{\rm var}$ between 3.7 and 6.5 per cent and excess variance detected at 3.6--8.5$\sigma$. Over the first and second halves of the observation separately the detections are weaker: in the second half the 0.6--0.8\,keV band is dominated by measurement noise, and the 1.9--3.0 and 3.0--7.0\,keV bands are only marginal in the first half. The source is therefore variable on timescales of a few to a few hundred ks. $F_{\rm var}$ increases with energy, from 3.8 per cent at 0.6--0.8\,keV to 6.5 per cent at 3.0--7.0\,keV, and a power-law fit gives $F_{\rm var}\propto E^{0.32\pm0.07}$ (Figure~\ref{fig:fvar}).

\begin{table}
\centering
\caption{Fractional variability of the SXT orbit-binned light curves over the full observation. $\sigma_{\rm NXS}$ is the uncertainty on the normalised excess variance.}
\label{tab:fvar}
\small
\setlength{\tabcolsep}{4pt}
\resizebox{\columnwidth}{!}{%
\begin{tabular}{lcccc}
\hline
Band (keV) & $N$ & $F_{\rm var}$ (per cent) & NXS & NXS$/\sigma_{\rm NXS}$ \\
\hline
0.6--0.8 & 48 & $3.8\pm0.5$ & 0.0014 & 3.6 \\
0.8--1.1 & 48 & $3.7\pm0.3$ & 0.0013 & 5.4 \\
1.1--1.4 & 48 & $4.8\pm0.3$ & 0.0023 & 7.1 \\
1.4--1.9 & 48 & $5.7\pm0.3$ & 0.0032 & 8.5 \\
1.9--3.0 & 48 & $5.4\pm0.5$ & 0.0030 & 5.7 \\
3.0--7.0 & 48 & $6.5\pm0.7$ & 0.0042 & 4.5 \\
\hline
\end{tabular}}
\end{table}

\subsection{Hardness ratios}

\begin{figure*}
\centering
\includegraphics[width=1.5\columnwidth]{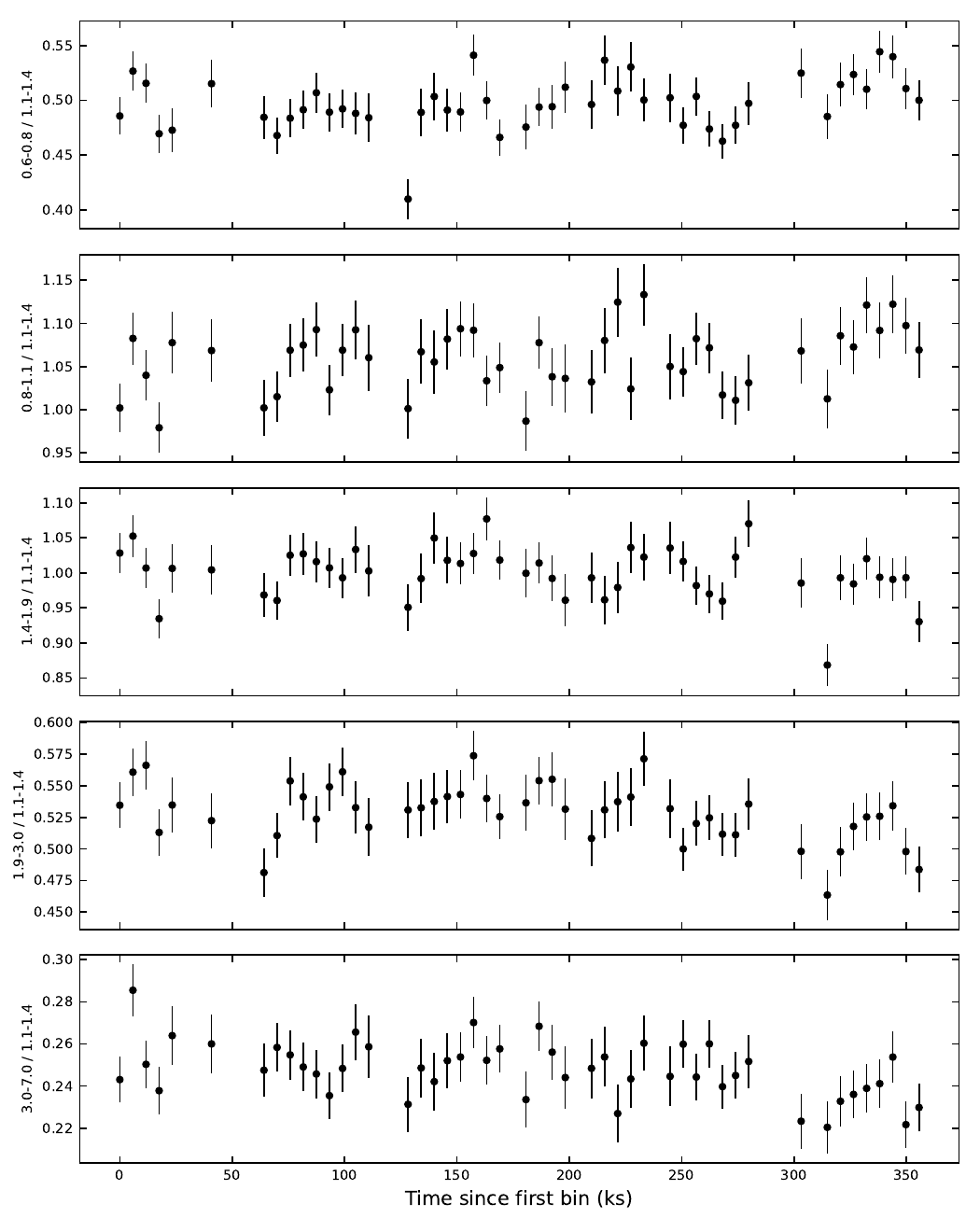}
\caption{Hardness ratios of each sub-band to the 1.1--1.4\,keV band as a function of time.}
\label{fig:hr}
\end{figure*}

Before the correlation analysis we checked whether the softest band (0.6--0.8\,keV) behaves differently from the others, since a drift in the low-energy response could mimic or hide a lag. We formed the ratios of each band to the 1.1--1.4\,keV band and tested them for excess scatter, linear trends in time, and rank correlations with time, with the flux in the remaining bands, and with the fractional exposure. The 0.6--0.8\,keV ratio shows no systematic trend that distinguishes it from the other ratios, and no ratio correlates with the fractional exposure (Figure~\ref{fig:hr}). One 0.6--0.8\,keV bin at 131\,ks is low: its ratio is 0.41 against a median of 0.495, $4.6\sigma$ below, and the bin is also visible in Figure~\ref{fig:lc}. Its fractional exposure (0.24) is among the lowest quarter of the bins (median 0.34), but its 691 counts are not unusual, and we found no instrumental reason for the dip. Leaving it out lowers $F_{\rm var}$ of the 0.6--0.8\,keV band from 3.8 to 3.1 per cent, within 1.4 times the quoted uncertainty, and changes the index of the $F_{\rm var}$--energy relation from 0.32 to about 0.34.

\subsection{Soft and hard band correlation and lags}

We cross-correlated one soft and one hard light curve with no overlap in energy, formed by summing the 0.6--0.8, 0.8--1.1 and 1.1--1.4\,keV bands (soft) and the 1.9--3.0 and 3.0--7.0\,keV bands (hard), in 1000\,s bins ($N=131$). Positive lags denote a hard lag (the hard band lags the soft band). The hard band has a large measurement noise contribution (intrinsic rms 0.054\,counts\,s$^{-1}$ against a mean error of 0.041\,counts\,s$^{-1}$), which limits the correlation coefficient that can be observed to about 0.7.

We computed the discrete correlation function \citep[DCCF;][]{Edelson1988} in 1000\,s delay bins out to $\pm20$\,ks, using only delays with at least a quarter of the number of data points in pairs. The peak and centroid lags and their uncertainties come from the flux randomisation and random subset selection method \citep{Peterson1998,Sun2018}, with 1000 realisations; the centroid uses the points above 80 per cent of the peak. To estimate how often such a correlation arises by chance, we simulated 1000 pairs of independent red-noise light curves \citep{Timmer1995} with the observed sampling, mean, variance and errors, and recorded the maximum DCCF over all delays in each pair. We used power-law PSDs with slopes of $-1.4$, $-1.7$ and $-2.0$, and the bending power laws of \citet{Chatterjee2018} derived for Mrk~421.

\begin{figure*}
\centering
\includegraphics[width=1.5\columnwidth]{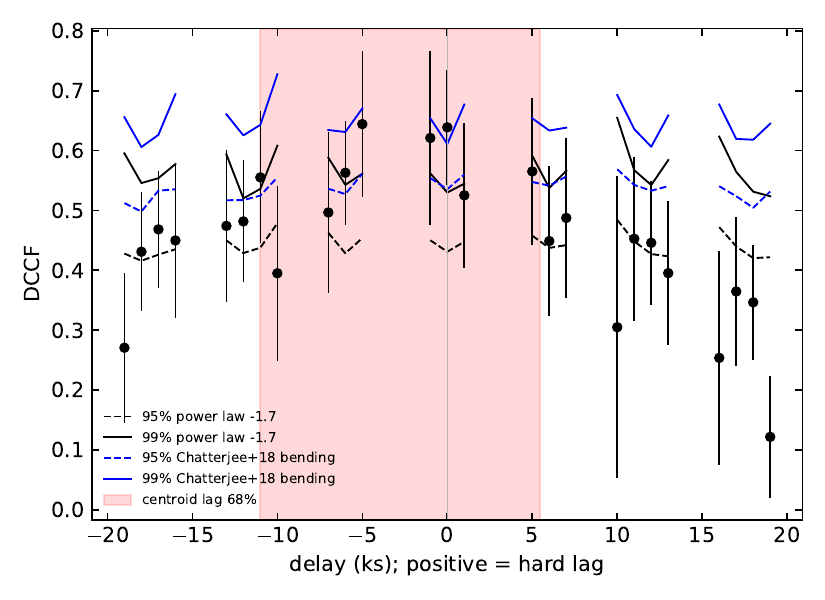}
\caption{DCCF between the soft (0.6--1.4\,keV) and hard (1.9--7.0\,keV) light curves. Dashed and solid lines show the 95 and 99 per cent levels reached at each delay by independent red-noise light curves with the observed sampling, for a power-law PSD of slope $-1.7$ (black) and for the bending power laws of \citet{Chatterjee2018} (blue). Delays with too few pairs are omitted. Positive delays mean the hard band lags the soft band. The shaded band is the 68 per cent range of the centroid lag.}
\label{fig:dccf}
\end{figure*}

\begin{figure*}
\centering
\includegraphics[width=1.5\columnwidth]{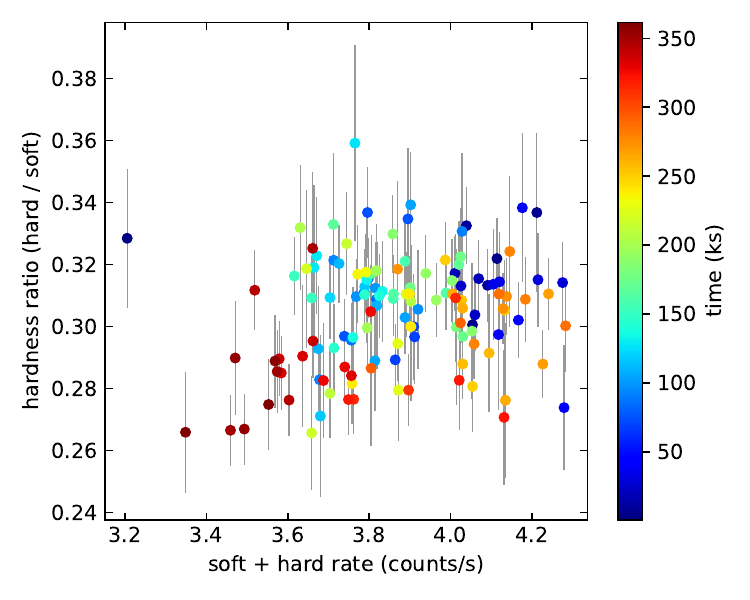}
\caption{Hardness ratio (1.9--7.0\,keV over 0.6--1.4\,keV) against total count rate, coloured by time.}
\label{fig:hid}
\end{figure*}

The DCCF peaks at $r=0.64$ at a delay of $-5$\,ks (Figure~\ref{fig:dccf}). The probability that independent red-noise light curves produce a peak this high at any delay is 0.015, 0.039 and 0.065 for slopes of $-1.4$, $-1.7$ and $-2.0$, and 0.096 for the bending power law. The correlation is therefore significant at about the 95 per cent level for shallow PSDs but not for steeper ones, and we do not regard it as established. The lag is consistent with zero: the peak lag is $-1^{+6}_{-10}$\,ks and the centroid lag is $-1.0^{+6.5}_{-10}$\,ks, so any lag between the two bands is smaller than about 6--10\,ks. These medians differ from the observed peak at $-5$\,ks because the DCCF is nearly flat between $-5$ and 0\,ks. The hardness ratio changes by about 10 per cent over the observation and shows no significant relation to the count rate (Spearman $\rho=0.16$, $p=0.06$, optimistic for red-noise data). The lowest-rate points at the end of the observation are slightly softer, and the track shows no loop (Figure~\ref{fig:hid}).

We also cross-correlated the 0.6--0.8\,keV band with each of the other five sub-bands, in 1000\,s and in 5830\,s bins. None of the five pairs shows a correlation that is significant after accounting for the number of trials (the smallest global $p$ of a single pair is 0.04 for 1000\,s bins and 0.03 for 5830\,s bins, for the shallowest PSD slope tried and larger for steeper slopes, which becomes about 0.2 and 0.16 after multiplying by the five pairs), and the 68 per cent interval of the centroid lag includes zero in every case, so any energy dependent lag is smaller than about 20\,ks. The single sub-band light curves have a large measurement noise contribution, so this test has little sensitivity.

\subsection{Power spectrum}

We estimated the power spectrum of the total 0.6--7\,keV light curve. Because the SXT sampling has large gaps, we used 1000\,s bins and a Lomb-Scargle periodogram with rms normalisation, evaluated at the Fourier frequencies $j/T$ from $2.8\times10^{-6}$\,Hz to just below the Nyquist frequency of $5\times10^{-4}$\,Hz. The Poisson noise level was fixed at the value expected from the measurement errors, $0.76$\,(rms/mean)$^2$\,Hz$^{-1}$.

The gaps put strong sidelobes at the orbital frequency (1.7$\times10^{-4}$\,Hz) and its harmonics, and a power-law model fitted directly to the periodogram returns a slope near $-0.6$ whatever the input slope is. We therefore modelled the continuum as $A\,L_\alpha(f) + C$, where $L_\alpha(f)$ is the mean periodogram of unit-rms power-law light curves with slope $\alpha$, generated with the method of \citet{Timmer1995} at 100\,s resolution and sampled exactly as the data (using the good time intervals of the event file), and $C$ is the Poisson level. The model parameters $A$ and $\alpha$ were fitted by minimising the Whittle statistic \citep{Vaughan2005}. Uncertainties on the slope come from a Neyman construction: for each input slope in a grid, we simulated 300 light curves with the observed mean, intrinsic variance and errors, applied the same estimator, and kept the input slopes for which the observed fitted slope falls within the central 68 (95) per cent of the simulated fitted slopes.

\begin{figure*}
\centering
\includegraphics[width=1.9\columnwidth]{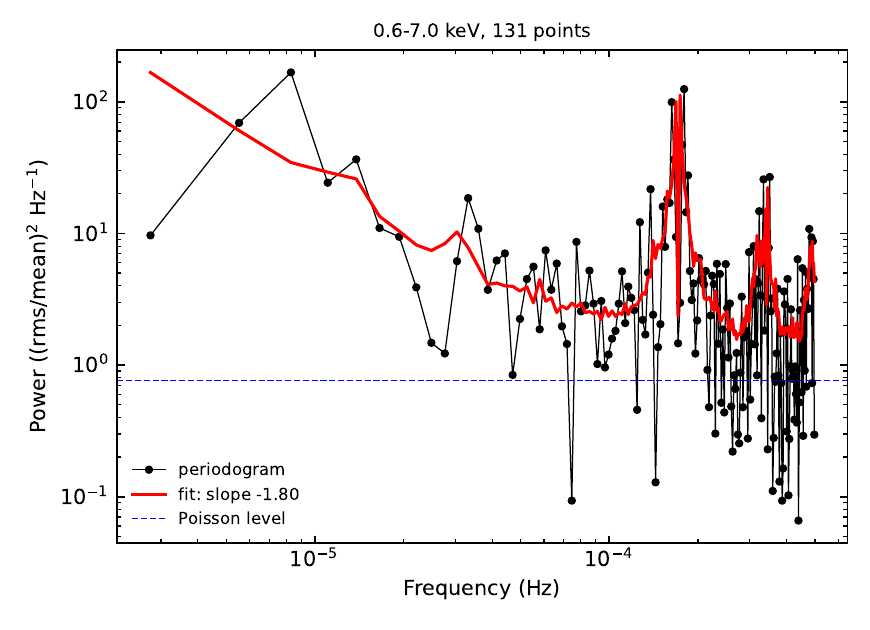}
\caption{Periodogram of the 0.6--7\,keV light curve in 1000\,s bins (points), the window-aware best-fitting model (red) and the Poisson level (dashed). The model includes the sidelobes produced by the orbital gaps.}
\label{fig:psd}
\end{figure*}

Figure~\ref{fig:psd} shows the periodogram and the best-fitting model. The best-fitting slope is $-1.80$, with a 68 per cent range of $-2.0$ to $-1.6$ and a 95 per cent range of $-2.8$ to $-1.4$ (the edges are limited to the 0.2 grid of input slopes). For input slopes steeper than $-2.0$ the fitted slope scatters over a wide range (for an input of $-2.0$ the 68 per cent range of fitted slopes is $-3.2$ to $-1.6$), so the slope is better constrained from the shallow side. The data are consistent with a power-law continuum: for input slopes of $-1.6$, $-2.0$ and $-2.4$ the fractions of simulations with a worse Whittle statistic are 0.61, 0.25 and 0.11, respectively. Above $10^{-4}$\,Hz the periodogram lies well above the Poisson level because low-frequency power leaks into high frequencies through the gaps, so the slope is set mainly by frequencies below $10^{-4}$\,Hz.

\subsection{Search for quasi-periodic oscillations}

We searched for periodicities with the Lomb-Scargle periodogram on a frequency grid oversampled by a factor of four, using the statistic $R(f) = 2I(f)/P(f)$, where $P(f)$ is the window-aware continuum above. The significance was derived by a parametric bootstrap \citep{Vaughan2010}: 1000 simulated light curves for each of four input slopes ($-1.2$, $-1.6$, $-2.0$ and $-2.4$) with the observed sampling, mean, variance and errors, each refitted with the same procedure, so that the uncertainty of the continuum is included. The global $p$ value is the fraction of simulations whose maximum $R$ over all frequencies exceeds the observed maximum, which includes the look-elsewhere effect. To set an upper limit on the amplitude of a periodic signal, we added sinusoids of known amplitude and random phase, sampled and averaged over the bins like the data, to simulated red-noise light curves with a slope of $-1.8$ and repeated the search (150 simulations per period and amplitude). A signal was counted as detected if $R$ near its frequency exceeded the 95th percentile of the maximum $R$ in simulations without a signal ($R=15.8$). The sinusoid was added to red noise with the observed variance, which makes the limit conservative.

\begin{figure*}
\centering
\includegraphics[width=\textwidth]{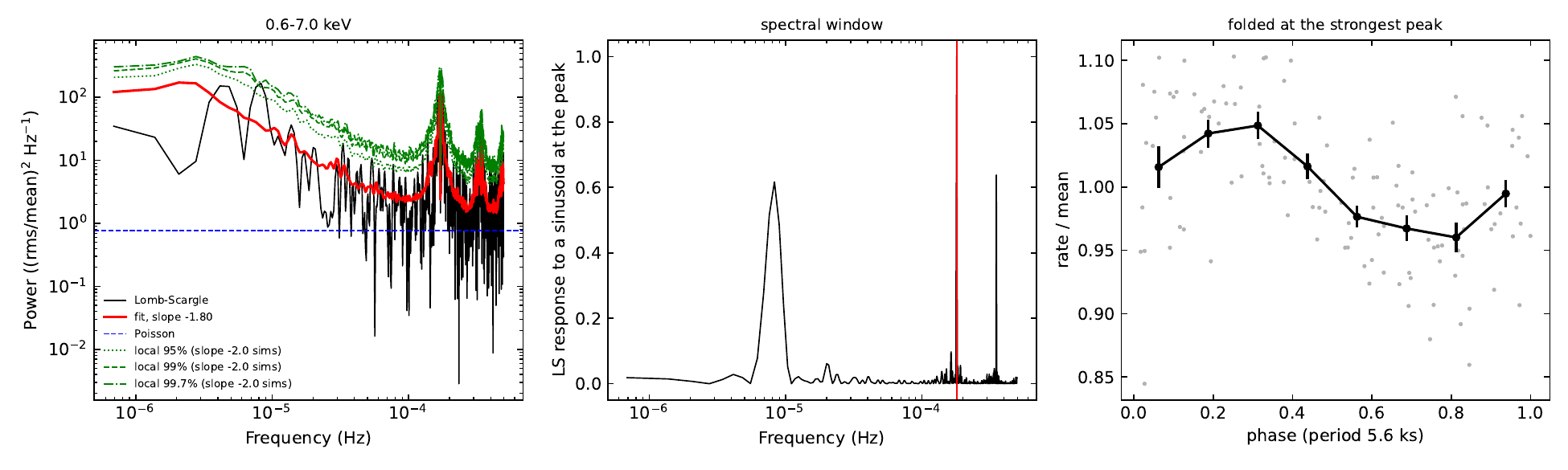}
\caption{Left: Lomb-Scargle periodogram with the window-aware fit and the local 95, 99 and 99.7 per cent levels from simulations. Middle: spectral window at the frequency of the strongest peak. Right: light curve folded at 5.6\,ks.}
\label{fig:qpo}
\end{figure*}

The highest peak is at 5.6\,ks ($R=10.5$), with comparable peaks at 3.0\,ks ($R=10.4$), 7.2\,ks ($R=9.9$) and 121\,ks ($R=9.7$). Its local $p$ is 0.002--0.004, but the global $p$ is 0.57--0.73 for the four input slopes, so the peak is not significant (Figure~\ref{fig:qpo} and Table~\ref{tab:psd}). The strongest two peaks are each within about $10^{-5}$\,Hz (three to four frequency resolution elements) of the orbital frequency and of its second harmonic, which may point to the window function but is not conclusive. The ranking of the highest peaks changed between runs with a smaller template library (200 simulations per slope, which is why the final run uses 1000), and the peaks have nearly equal $R$, so none of them stands out. The feature at 121\,ks covers only three cycles in the data, well below the roughly five cycles needed to separate a periodicity from red noise \citep{Vaughan2016}. We find no evidence for a QPO in the SXT light curve. A sinusoid with an rms amplitude of 1.9 to 3.7 per cent of the mean rate would have been detected with 90 per cent probability at any period between 3 and 50\,ks (1.9 per cent at 4\,ks, 2.0 per cent at 15\,ks and 3.7 per cent at 50\,ks), which is 13 to 51 per cent of the intrinsic variance of the light curve. The sensitivity is lower at 3.0 and 5.0\,ks (3.5 and 3.1 per cent), which may reflect the orbital window. It falls at longer periods: 5.8 per cent at 80\,ks (4.5 cycles) and above 10 per cent at 120\,ks (3 cycles). Folded at 5.6\,ks (Figure~\ref{fig:qpo}, right panel), the light curve shows a smooth modulation in eight phase bins, with a maximum of 1.05 times the mean near phase 0.3 and a minimum of 0.96 near phase 0.8. The peak-to-peak amplitude is about 9 per cent, which is an rms of about 3 per cent for a sinusoid and is close to the detection limit of 3.1 per cent found at 5.0\,ks. The individual points scatter widely about the profile, and the uncertainty of each phase bin is 1 to 1.5 per cent. Because 5.6\,ks is close to the orbital period of 5.83\,ks, the profile is shaped by the sampling as well as by the source, and it is not independent evidence for a periodic signal.

\begin{table}
\centering
\caption{Summary of the PSD and QPO results for the 0.6--7\,keV light curve in 1000\,s bins.}
\label{tab:psd}
\small
\begin{tabular}{p{0.68\columnwidth}c}
\hline
Quantity & Value \\
\hline
Fitted PSD slope & $-1.80$ \\
68 per cent range & $-2.0$ to $-1.6$ \\
95 per cent range & $-2.8$ to $-1.4$ \\
Strongest peak (period) & 5.6\,ks \\
Global $p$ (input slopes $-1.2$ to $-2.4$) & 0.57--0.73 \\
rms amplitude detectable at 90 per cent, 3--50\,ks & 1.9--3.7 per cent \\
\hline
\end{tabular}
\end{table}

\subsection{Flux distribution}

\begin{figure*}
\centering
\includegraphics[width=1.5\columnwidth]{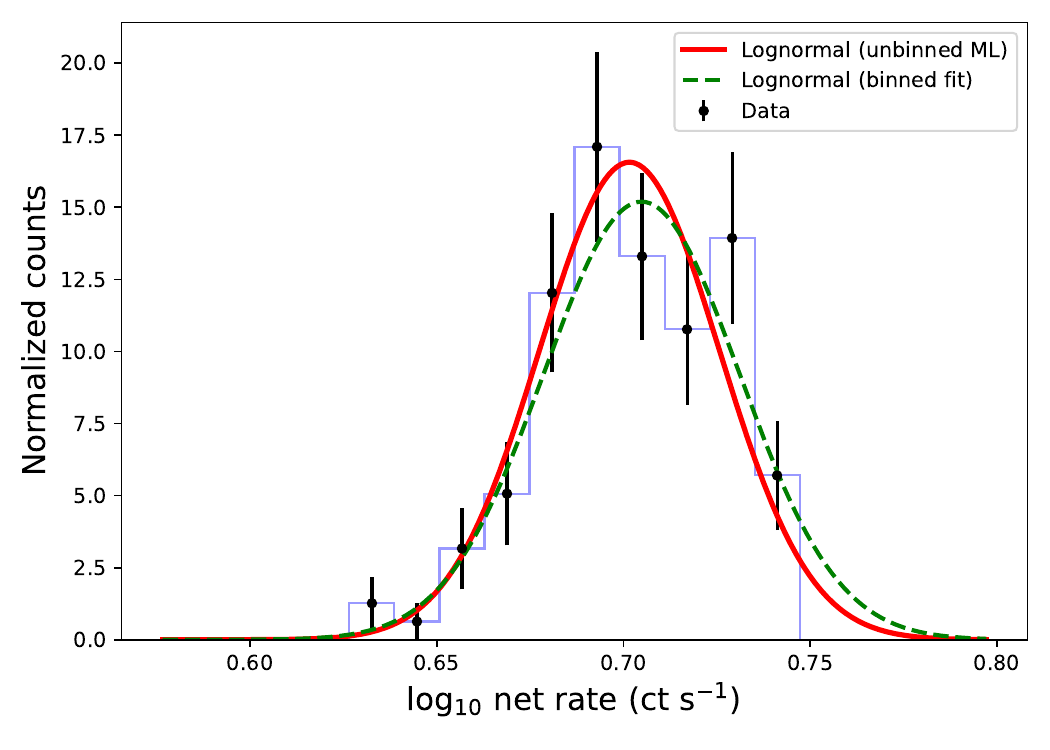}
\caption{Distribution of $\log_{10}$ of the 0.6--7\,keV count rate in 1000\,s bins, with the lognormal fit (maximum likelihood, solid; histogram fit, dashed).}
\label{fig:flux}
\end{figure*}

We examined the distribution of the 0.6--7\,keV count rate in 1000\,s bins ($N=131$). It is described by a lognormal with a median of $5.03$\,counts\,s$^{-1}$ and a width of $0.0241\pm0.0014$\,dex in $\log_{10}F$ (maximum likelihood, with the error from a bootstrap), corresponding to a scatter of 5.5 per cent in flux (Figure~\ref{fig:flux}). After subtracting the mean measurement error (0.0081\,dex) in quadrature, the intrinsic width is $0.0227$\,dex ($\sigma_{\ln}=0.052$). A Kolmogorov--Smirnov test does not reject the lognormal ($p=0.79$), and a fit to the histogram (10 bins) gives a consistent centre of $0.7048\pm0.0029$\,dex, a width of $0.0263\pm0.0023$\,dex and $\chi^2/{\rm dof}=6.95/7$. The uncertainties are lower limits, because neighbouring points of a red-noise light curve are correlated.

\subsection{Light curve profile}

To describe the shape of the 0.6--7\,keV light curve in 5830\,s bins we fitted a constant plus a sum of exponential peaks, each with a rise time $\tau_r$ and a decay time $\tau_d$ (e-folding times), with up to three peaks. Each fit was started from the local maxima of the smoothed light curve and from 30 randomly perturbed starting points. The lower limit on $\tau_r$ and $\tau_d$ is one orbit bin (5830\,s), because faster changes cannot be resolved. A peak counts as well constrained if every parameter exceeds twice its uncertainty and none lies at a bound; uncertainties are scaled by $\sqrt{\chi^2_{\rm red}}$ when $\chi^2_{\rm red}>1$.

\begin{figure*}
\centering
\includegraphics[width=1.5\columnwidth]{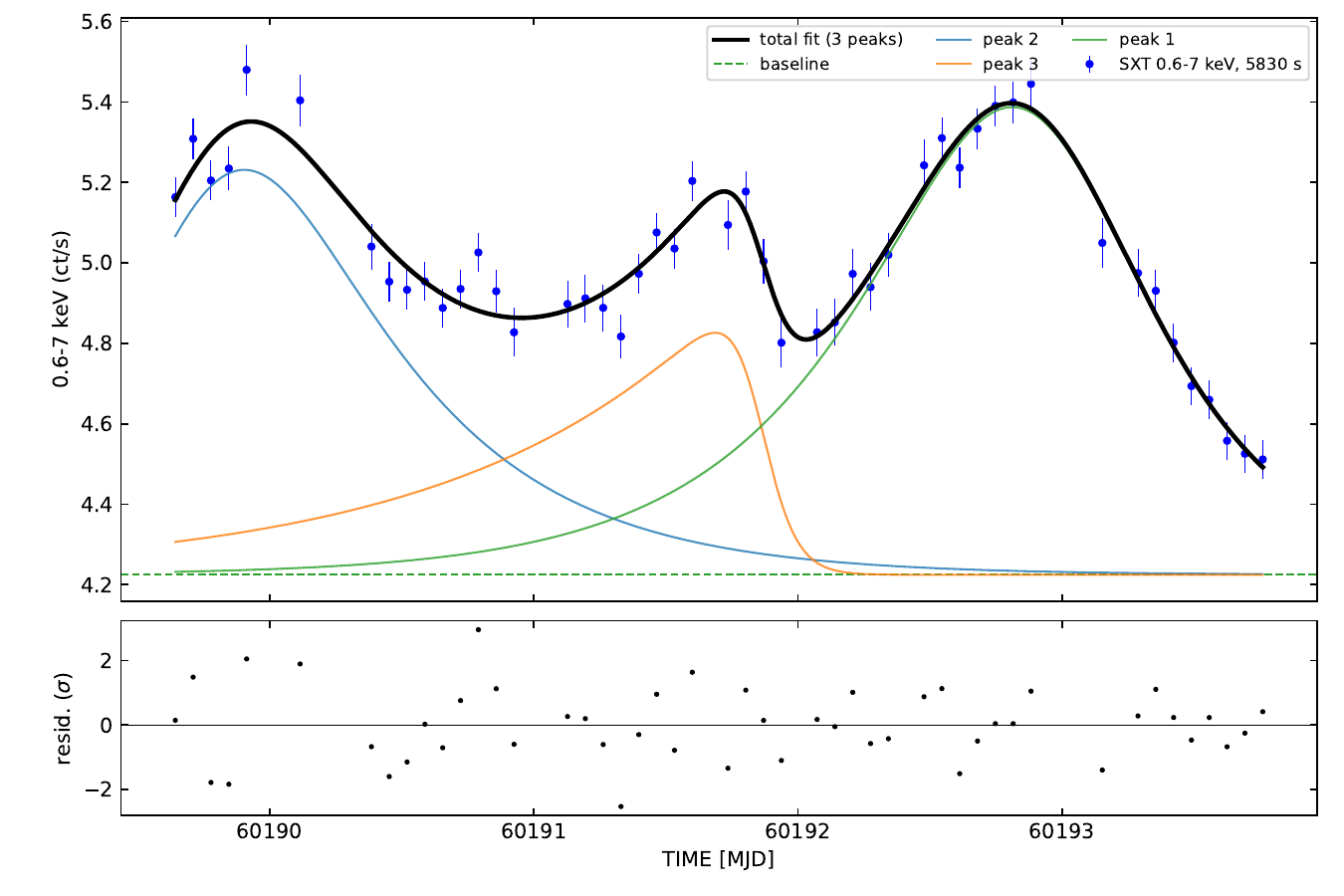}
\caption{SXT 0.6--7\,keV light curve in 5830\,s bins with the best-fitting model (baseline plus three exponential peaks) and its components. The lower panel shows the residuals.}
\label{fig:flare}
\end{figure*}

The Bayesian information criterion selects three peaks ($\chi^2=59.5$ for 48 points and 13 parameters, $\chi^2_{\rm red}=1.70$; Figure~\ref{fig:flare}). Only one of the three peaks is well constrained: a hump at MJD $60192.89\pm0.15$ with a peak amplitude $F_0=1.15\pm0.22$\,counts\,s$^{-1}$, $\tau_r=0.57\pm0.14$\,d and $\tau_d=0.41\pm0.16$\,d (Table~\ref{tab:hump}). The profile is nearly symmetric, $\zeta=(\tau_d-\tau_r)/(\tau_d+\tau_r)=-0.16$, and the corresponding doubling and halving times are 34 and 24\,ks. A second peak lies at the start of the observation, where the data do not cover its rise, and the decay time of a third is at the lower bound of one orbit bin, so neither is individually constrained. Removing any one of the three raises $\chi^2$ by more than 100, so all three are needed to describe the shape, but the parameters of the two poorly constrained peaks are descriptive only.

\begin{table}
\centering
\caption{Parameters of the well-constrained hump in the 0.6--7\,keV light curve (5830\,s bins). $t_0$ is the time of maximum, $F_0$ the peak amplitude above the other components, and $\tau_r$, $\tau_d$ the rise and decay e-folding times.}
\label{tab:hump}
\begin{tabular}{lc}
\hline
Parameter & Value \\
\hline
$t_0$ (MJD) & $60192.89\pm0.15$ \\
$F_0$ (counts\,s$^{-1}$) & $1.15\pm0.22$ \\
$\tau_r$ (d) & $0.57\pm0.14$ \\
$\tau_d$ (d) & $0.41\pm0.16$ \\
$|\zeta|$ & 0.16 \\
\hline
\end{tabular}
\end{table}

Exponential peaks can describe a red-noise light curve well even when it contains no distinct flares, so we tested the result with simulations. We generated 300 red-noise light curves for each of three PSD slopes ($-1.4$, $-1.7$ and $-2.0$) with the observed sampling, mean, intrinsic variance and errors, and fitted them in the same way. The BIC selects three or more peaks in 95, 88 and 76 per cent of the simulations, and the improvement in BIC over a constant is as large as observed in 9, 17 and 23 per cent. A multi-peak decomposition is therefore not significant. At least one well-constrained peak, as defined above, appears in 1.0, 4.7 and 7.0 per cent of the simulations, so the hump near MJD 60192.9 is marginally distinguishable from red noise ($p=0.05$ to 0.07 for slopes of $-1.7$ to $-2.0$, about 2$\sigma$). For the shallowest slope, which is still inside the 95 per cent range of the fitted slope, $p=0.01$, so the result depends on the continuum slope.

\section{Discussion}
\label{sec:disc}

\subsection{Variability and spectral behaviour}

All six sub-bands vary at the 3.7 to 6.5 per cent level in $F_{\rm var}$, which is lower than the 9 to 15 per cent that \citet{Markowitz2022} measured in the X-ray band of Mkn 421 during a 4\,d campaign. They found that $F_{\rm var}$ increases monotonically with energy, as $F_{\rm var}\propto E^{0.26\pm0.02}$. Over the SXT range $F_{\rm var}$ also rises with energy, with an index of $0.32\pm0.07$, consistent with that value. In the October 2017 flaring state of 1ES 1959+650, when the SXT rate was about 14\,counts\,s$^{-1}$, \citet{Shah2021} measured $F_{\rm var}=8.4\pm1.2$ per cent in the 0.7--7\,keV band in 5\,ks bins, which is higher than our values at a rate of about 5\,counts\,s$^{-1}$.
The hardness ratio of the 1.9--7\,keV band to the 0.6--1.4\,keV band changes by about 10 per cent and has no significant relation to the count rate ($\rho=0.16$, $p=0.06$, which is optimistic for red-noise data). \citet{DasChatterjee2023} found a harder-when-brighter trend in all epochs of 1ES 1959+650 using hardness ratios between 7--20\,keV and 0.7--7\,keV. The SXT bands cover a much smaller range of energy and the signal in the hard band is small, so we cannot test that trend with the same sensitivity. The lowest-rate points at the end of the observation are slightly softer, and the track in the hardness-intensity plane shows no loop.

\subsection{Correlation and lags}

The soft and hard light curves are correlated near zero lag, but at about the 95 per cent level and only for shallow power spectra. For a slope of $-1.7$ the chance probability of the peak is 0.04, and it rises to 0.07 for a slope of $-2.0$ and to 0.10 for the bending power law of \citet{Chatterjee2018}. This matters because the significance of a peak at a single delay is higher than the probability that a peak this high occurs at any delay. The observed peak lies above the 99 per cent line at its own delay but has a global probability of a few per cent. The hard band has a large measurement noise contribution, which limits the observable correlation coefficient to about 0.7 and leaves the test with little sensitivity.

The lag is consistent with zero and is limited to about 10\,ks. The lags that \citet{DasChatterjee2023} measured for 1ES 1959+650 are between 0.3 and 7\,ks. These lie within about $1\sigma$ of our centroid lag, so our data neither confirm nor rule them out. In the same way, the lead of the 0.6--0.8\,keV band of $4.6\pm2.6$\,ks found by \citet{Markowitz2022} for Mkn 421 cannot be tested. The centroid lags of the pairs that include the 0.6--0.8\,keV band are mostly negative (the higher-energy band leads), between $-8$ and $+3$\,ks, with uncertainties of 6 to 18\,ks and no pair significant after allowing for five trials. The sign is opposite to a lead of the 0.6--0.8\,keV band, but every interval includes zero, so the data neither support nor contradict that lead. Our light curves cover a single orbit-gapped observation with 131 bins, whereas an energy-dependent lag of a few ks would need a longer continuous exposure or a brighter hard band.

Because no lag is detected, this observation gives no estimate of the magnetic field strength or the acceleration parameter.

\subsection{Power spectrum and periodicity}

The slope of $-1.80$ (95 per cent range $-2.8$ to $-1.4$) lies between the low- and high-frequency slopes of the bending power law that \citet{Chatterjee2018} fitted to the X-ray PSD of Mkn 421 ($-1.1$ to $-1.4$ below the break and about $-2.5$ above it). They place the break near $10^{-6}$ to $2\times10^{-6}$\,Hz in the 2--10 and 0.3--2\,keV bands, each uncertain by about one dex. Our lowest Fourier frequency is $2.8\times10^{-6}$\,Hz, so a break in this range would sit at the edge of our data. A single power law fits the periodogram acceptably, and we cannot say whether the PSD of 1ES 1959+650 bends. The fitted slope scatters widely for input slopes steeper than $-2.0$, so the data constrain the slope mainly from the shallow side. The slope is also consistent with the $-1.96\pm0.14$ that \citet{Markowitz2022} fitted to the periodogram of the 1.4--1.9\,keV SXT light curve of Mkn~421.

The search for periodicities finds no significant peak. The strongest, at 5.6\,ks, has a global $p$ of 0.57 to 0.73, and the peaks at 3.0, 7.2 and 121\,ks have almost the same $R$. A peak at the 121\,ks timescale covers only three cycles. The upper limit on a sinusoidal signal is 1.9 to 3.7 per cent in rms amplitude between 3 and 50\,ks, so the data exclude an oscillation that carries more than 13 to 51 per cent of the intrinsic variance in that range. The limit rises quickly above 50\,ks, where fewer than seven cycles are covered, and the search is not sensitive at 120\,ks. It was derived for a single continuum slope ($-1.8$) and a coarse amplitude grid, so it is approximate.

\subsection{Flux distribution}

The count rate distribution is consistent with a lognormal of width 0.024\,dex (Kolmogorov--Smirnov $p=0.79$), which corresponds to a scatter of about 5 per cent in flux. A lognormal distribution is often taken to indicate that the variability is multiplicative. Our result is consistent with that picture, but the rms is small and the 131 points are correlated, so a single epoch cannot establish it. Observations with larger variability, or a longer series with several flux states, are needed for that.

\subsection{Light curve profile}

The light curve is described by one well-constrained hump near MJD 60192.9 with rise and decay times of $0.57\pm0.14$ and $0.41\pm0.16$\,d, so the amplitude of the hump doubles or halves in 24 to 34\,ks (the total flux changes by a much smaller fraction, because the hump sits on a baseline). The profile is nearly symmetric. These times are 2.4 to 3.4 times longer than the lag limit of about 10\,ks from the cross-correlation, so the two results do not conflict, but the lag limit also does not reach the shorter timescales. The reduced $\chi^2$ of 1.7 for the preferred three-peak model shows that structure remains on shorter timescales than the fit describes. The simulations show that red noise alone gives a three-peak solution in most trials, and a well-constrained peak in 1 to 7 per cent of them, so we regard the hump as marginal and do not interpret the shapes of the other two peaks.

\section{Summary}
\label{sec:summary}

We analysed the AstroSat/SXT observation of 1ES 1959+650 of 2023 September 2 and obtain the following results.

\begin{itemize}
\item All six sub-bands between 0.6 and 7\,keV show excess variance, with fractional variability of 3.7 to 6.5 per cent over 362\,ks. It increases with energy, $F_{\rm var}\propto E^{0.32\pm0.07}$.
\item The 0.6--7\,keV power spectrum is a red-noise power law with slope $-1.80$ (95 per cent range $-2.8$ to $-1.4$), fitted with a model that includes the window function of the orbital gaps. The slope is constrained mainly from the shallow side. A break, such as the one reported for Mkn 421, cannot be tested.
\item No periodicity is significant: the strongest peak, at 5.6\,ks, has a global $p$ of 0.57 to 0.73. A sinusoid with an rms amplitude above 1.9 to 3.7 per cent between 3 and 50\,ks would have been detected with 90 per cent probability.
\item The soft (0.6--1.4\,keV) and hard (1.9--7\,keV) light curves are correlated near zero lag only marginally (global $p=0.04$ and 0.07 for power-spectrum slopes of $-1.7$ and $-2.0$). The lag is $-1^{+6}_{-10}$\,ks, consistent with zero.
\item Pairs of sub-bands that include 0.6--0.8\,keV show no significant correlation or lag after allowing for the number of trials, so a lead of the 0.6--0.8\,keV band of a few ks can be neither confirmed nor excluded.
\item The light curve has one hump near MJD 60192.9 with rise and decay times of $0.57\pm0.14$ and $0.41\pm0.16$\,d. It is only marginally distinguishable from red noise ($p\approx0.05$ to $0.07$ for slopes of $-1.7$ to $-2.0$), and a decomposition into several peaks is not significant.
\item The count rate distribution (1000\,s bins) is consistent with a lognormal of width $0.0241\pm0.0014$\,dex.
\item The hardness ratio has no significant dependence on flux, and the hardness-intensity track shows no loop.
\end{itemize}

\section*{Data availability}
The AstroSat data are public through the ISSDC archive.

\section*{Acknowledgements}
SA and ZS  express  gratitude to the Inter-University Centre for Astronomy and Astrophysics (IUCAA) in Pune, India, for the support and facilities provided.

\bibliographystyle{elsarticle-harv} 
\bibliography{references}






\end{document}